\documentclass[sigconf, nonacm]{acmart}

\usepackage{algorithmic}
\usepackage{textcomp}
\usepackage{subfig}
\usepackage{multirow}
\usepackage{enumitem}
\usepackage{pifont}

\begin{document}

\title{Detecting Adversarial Evasion Attacks Against Autoencoder-Based Network Intrusion Detection Systems}

\author{Niklas Bunzel}
\affiliation{%
  \institution{Fraunhofer SIT / TU Darmstadt / ATHENE}
  \city{Darmstadt}
  \country{Germany}
}
\email{niklas.bunzel@sit.fraunhofer.de}

\author{Ashim Siwakoti}
\affiliation{%
  \institution{TU Darmstadt}
  \city{Darmstadt}
  \country{Germany}
}
\email{siwakoti.ashim1995@gmail.com}

\begin{abstract}
	Evasion attacks deliberately manipulate input to an ML-based system to produce an incorrect prediction while the manipulated input still appears benign. The PANDA framework has demonstrated that adversarial examples developed for the vision domain can be transferred to the network domain by converting packet sequences into invertible grayscale images, enabling gradient-based attacks such as masked FGSM against autoencoder-based network intrusion detection systems (NIDS). These attacks manipulate the NIDS anomaly score without altering the underlying attack semantics, leaving defenders without a straightforward way to distinguish between benign flows and carefully perturbed malicious traffic. In this paper, we propose two complementary detectors: the \emph{Residual Localisation Detector (RLD)}, which tracks the spatial concentration of reconstruction errors in the inter-arrival time feature region in image space; and the \emph{Feature-Space Perturbation Consistency (FPC) Detector}, which operates directly on packet-level inter-arrival time features in packet-feature space. We evaluate both detectors on benign, malicious, and adversarial traffic from multiple IoT devices in the UQ-IoT dataset. Both detectors achieve near-perfect detection performance (TNR, TPR, precision, recall, and F1-score $\geq 0.99$) against adversarial examples across the evaluated IoT traffic. Our results indicate that integrating reconstruction-based scoring with perturbation consistency checks, in both image space and packet-feature space, offers a practical defence against emerging PANDA-style adversarial attacks on NIDS.
\end{abstract}

\maketitle


\section{Introduction}

Networked systems are routinely exposed to a broad range of attacks, from denial-of-service and data exfiltration to lateral movement across enterprise and IoT environments~\cite{Li_2022,ennaji2024adversarialchallengesnetworkintrusion,Biggio_2018}. Network Intrusion Detection Systems (NIDS) play a central role in such environments by observing traffic and raising alarms when flows deviate from expected behaviour~\cite{Li_2022,mirsky2018kitsuneensembleautoencodersonline}. Classical NIDS either rely on hand-crafted signatures that match known attack patterns, or on anomaly-based techniques that learn a model of benign traffic and flag deviations~\cite{5504793,ALDWEESH2020105124}. In recent years, anomaly-based NIDS have increasingly adopted machine-learning and deep-learning models, as these can capture complex non-linear relationships and often achieve higher detection rates~\cite{ennaji2024adversarialchallengesnetworkintrusion,10005100}.

In parallel, adversarial machine learning (AML) has shown that even highly accurate ML models can be forced into misclassification by small, carefully crafted perturbations~\cite{goodfellow2015explainingharnessingadversarialexamples,szegedy2014intriguingpropertiesneuralnetworks}. This phenomenon was first studied in computer vision, but has since been applied to security-relevant tasks including NIDS~\cite{Biggio_2018,10.1145/3339252.3339266}. Transferring adversarial techniques to NIDS is more challenging than to image classifiers because of the mismatch between \emph{feature space} and \emph{problem space}~\cite{cortellazzi2024intriguingpropertiesadversarialml,swain2024panda}. In image classification, the model operates directly on pixel values and gradients can be computed with respect to each pixel. In NIDS, raw packets must first be parsed and transformed into higher-level features, and these feature extractors are typically non-differentiable~\cite{practicalTSAA}.

The PANDA framework by Swain et al.\ addresses this gap by proposing an invertible representation that enables gradient-based adversarial attacks directly on network traffic~\cite{swain2024panda}. PANDA converts a sequence of packets into a grayscale image by extracting selected header and timing fields. A convolutional autoencoder (CNN-AE) is trained as an anomaly-based surrogate NIDS on these images. Because the mapping from packets to image is invertible, adversarial images generated against the surrogate can be translated back into valid packet sequences. PANDA uses a masked variant of the Fast Gradient Sign Method (FGSM) that perturbs only the bits corresponding to inter-arrival time (IAT), preserving packet semantics while enabling efficient gradient-based optimisation. On the UQ-IoT dataset~\cite{He2022_UQIoTIDS2021}, such adversarial traces reliably evade the surrogate autoencoder and transfer to other NIDS models as well.

These results suggest that once an invertible, differentiable representation is available, much of the adversarial toolbox from computer vision becomes applicable to NIDS. However, most current NIDS deployments follow a simple pattern: the system computes an anomaly score and compares it to a fixed threshold, without explicitly checking whether this score may have been manipulated by an adversary. If an attacker can use a PANDA-style pipeline to adjust timing features just enough to cross below the threshold, the attack may be reclassified as benign while retaining its original impact.

This paper starts from the assumption that such attacks are not purely theoretical and asks: \emph{Given an autoencoder-based NIDS under PANDA-style attack, can we design practical detectors that reveal adversarial network traffic, even when it appears benign to the underlying NIDS?}

To explore this question, we adopt PANDA's surrogate model and packet-to-image representation as our attack backbone and focus on detector design and evaluation. We propose two complementary detectors: the \emph{Residual Localisation Detector (RLD)}, which tracks the spatial concentration of reconstruction errors; and the \emph{Feature-Space Perturbation Consistency (FPC) Detector}, which operates directly on packet-level IAT features.

The main contributions of this paper are:
\begin{itemize}
    \item We design two adversarial detectors tailored to PANDA-style attacks on autoencoder-based NIDS; the RLD and FPC detectors each exploit different aspects of reconstruction behaviour and perturbation consistency in image space or packet-feature space.
    \item We evaluate the performance, strengths, and limitations of the detectors on benign and adversarial traffic from multiple IoT devices and attack types, demonstrating near-perfect detection across both methods.
\end{itemize}

\section{Related Work}

\subsection{Network Intrusion Detection Systems}

Traditional NIDS rely on signature-based detection, comparing observed network data to a database of known attack patterns~\cite{5504793,ALDWEESH2020105124}. While effective for known threats with low false alarm rates, they fail to detect zero-day or polymorphic variants until new signatures are created~\cite{ALDWEESH2020105124,LIAO201316}. Anomaly-based NIDS address this by learning a model of normal network behaviour and flagging deviations, enabling detection of previously unseen attacks~\cite{ALDWEESH2020105124,s21134294}.

Autoencoder-based methods have recently gained significant attention because they can model complex, high-dimensional network traffic without labelled attack examples~\cite{swain2024panda,chen2018autoencoder}. When trained on benign traffic, the autoencoder learns to reconstruct normal patterns with low error, while anomalous traffic results in higher reconstruction error~\cite{swain2024panda,s21134294}. Systems like Kitsune use ensembles of autoencoders for real-time anomaly detection~\cite{mirsky2018kitsuneensembleautoencodersonline}. However, recent studies show that autoencoder-based NIDS can be vulnerable to adversarial manipulation~\cite{ARAFAH2025112455,zhang2024toward,swain2024panda}.

\subsection{Adversarial Machine Learning}

Adversarial examples are inputs designed by introducing small perturbations to legitimate data, causing the model to produce incorrect predictions~\cite{szegedy2014intriguingpropertiesneuralnetworks}. Adversarial attacks can be classified by the attacker's objective: evasion attacks alter input during inference, while poisoning attacks target the training process~\cite{Biggio_2018}. A key dependency is the differentiability of the target model, as many attacks exploit gradient information~\cite{demontis2019adversarialattackstransferexplaining}.

The Fast Gradient Sign Method (FGSM), introduced by Goodfellow et al.~\cite{goodfellow2015explainingharnessingadversarialexamples}, is a widely cited gradient-based attack:
\begin{equation}
x_{\text{adv}} = x + \epsilon \cdot \text{sign}(\nabla_x J(\theta, x, y))
\label{eq:fgsm}
\end{equation}
where $\epsilon$ controls the perturbation magnitude. Despite its simplicity, FGSM remains a standard baseline and building block for more advanced attacks~\cite{goodfellow2015explainingharnessingadversarialexamples,kurakin2017adversarialexamplesphysicalworld}.

\subsection{Adversarial Attacks on NIDS}

ML-based NIDS have become targets of adversarial attacks due to their widespread deployment~\cite{zhang2024toward,10005100}. Unlike vision-based attacks where perceptual similarity is the primary concern, adversarial perturbations on NIDS must also follow protocol constraints and preserve the malicious functionality~\cite{elshehaby2025sokadversarialevasionattacks,KUMAR2025126513}. Prior work operates either in feature space, perturbing feature vectors directly~\cite{practicalTSAA,Lin_2022}, or in problem space, producing valid, replayable adversarial traffic~\cite{swain2024panda,10.1145/3359992.3366642}. Feature-space attacks demonstrate theoretical vulnerabilities but often assume modified feature vectors correspond to feasible network traffic, which may not hold in practice~\cite{practicalTSAA,cortellazzi2024intriguingpropertiesadversarialml}.

\section{PANDA Attack Framework}

This section describes the PANDA framework~\cite{swain2024panda}, which serves as the attack backbone for our work.

\subsection{Packet-to-Image Representation}

Unlike image data, network packets do not naturally allow for gradient-based perturbations because they live in a discrete problem space. PANDA overcomes this by converting a sequence of raw packets into a single grayscale image. For each packet, it extracts: inter-arrival time (IAT), source/destination IP and MAC addresses, source/destination port numbers, and frame length. These fields are converted into binary values to form a 235-bit vector per packet. By stacking 235 such vectors, PANDA creates a $235 \times 235$ grayscale image where each row represents one packet and each column represents one bit position across all packets. This representation is \emph{invertible}: each row can be decoded to reconstruct the original packet sequence.

\subsection{Surrogate CNN Autoencoder}

PANDA trains a convolutional autoencoder (CNN-AE) as a surrogate NIDS. The CNN-AE is trained strictly on benign traffic to minimise reconstruction loss. At inference time, the anomaly score is defined as the negative reconstruction error:
\begin{equation}
a(x) = -\, L(x, \hat{x})
\end{equation}
where $\hat{x}$ is the autoencoder reconstruction. A decision threshold $\lambda$ is set at the 95th percentile of reconstruction errors on benign training data. Samples with $a(x) \geq \lambda$ are classified as benign; others as malicious.

\subsection{Masked FGSM}

PANDA uses a masked variant of FGSM to generate adversarial examples. A binary mask $M \in \{0,1\}^{H \times W}$ restricts perturbations to the IAT bit positions (first 32 columns), leaving IP/MAC addresses, ports, and frame length untouched:
\begin{equation}
x_{\text{adv}} = \Pi_{[0,1]}\Bigl(x_0 + \epsilon \cdot \operatorname{sign}\!\bigl(\nabla_{x_0} L(f_\theta(x_0), x_0)\bigr) \odot M\Bigr)
\label{eq:masked_fgsm}
\end{equation}
where $\Pi_{[0,1]}$ clips to $[0,1]$ and $\odot$ is the element-wise product. By concentrating on timing features, PANDA preserves packet validity while evading detection, making it a practical grey-box attack.

\section{System and Threat Model}

\subsection{Dataset and Environment}

All experiments use the UQ-IoT-IDS-2021 dataset~\cite{He2022_UQIoTIDS2021}, which captures network traffic from genuine IoT devices (smartphones, smart TVs, IP cameras, smart speakers) connected to a common wireless network. The dataset contains benign traffic recorded over one week and nine attack types: host discovery, port scanning, service detection, ARP spoofing, Telnet brute force, SYN/UDP/HTTP/ACK flooding. Following the PANDA setup, we focus on service detection, port scanning, and ARP spoofing, plus benign traffic.

Packet pre-processing follows PANDA's pipeline: each packet is parsed into a 235-bit vector (IAT, MAC, IP, ports, frame length), and sequences of 235 packets are stacked into $235 \times 235$ grayscale images. Only benign traffic trains the surrogate autoencoder; both benign and malicious traffic are used for detector evaluation.

\subsection{Attacker Model}

The threat model follows PANDA's assumptions~\cite{swain2024panda}. The attacker's objectives are: (1) preserve the malicious effect of the original attack, and (2) evade the NIDS by pushing the anomaly score above the detection threshold. The attacker operates in a grey-box setting: they know the model type and input features, can collect benign IoT traffic, and can train a surrogate NIDS. We assume the attacker is unaware of the proposed detectors.

\subsection{Defender Perspective}

The baseline NIDS is exposed to both clean and PANDA-modified malicious traffic. The proposed detectors are introduced as additional modules that observe the same input representation and optionally the NIDS output, flagging sequences exhibiting characteristics of PANDA-style perturbations.

\section{Proposed Detection Methods}

We propose two detectors that exploit structural inconsistencies introduced by gradient-based manipulation. RLD operates in PANDA's image space, while FPC operates directly on packet-level features.

\subsection{Residual Localisation Detector (RLD)}

\begin{figure}[t]
    \centering
    \includegraphics[width=0.35\textwidth,height=0.65\textheight,keepaspectratio]{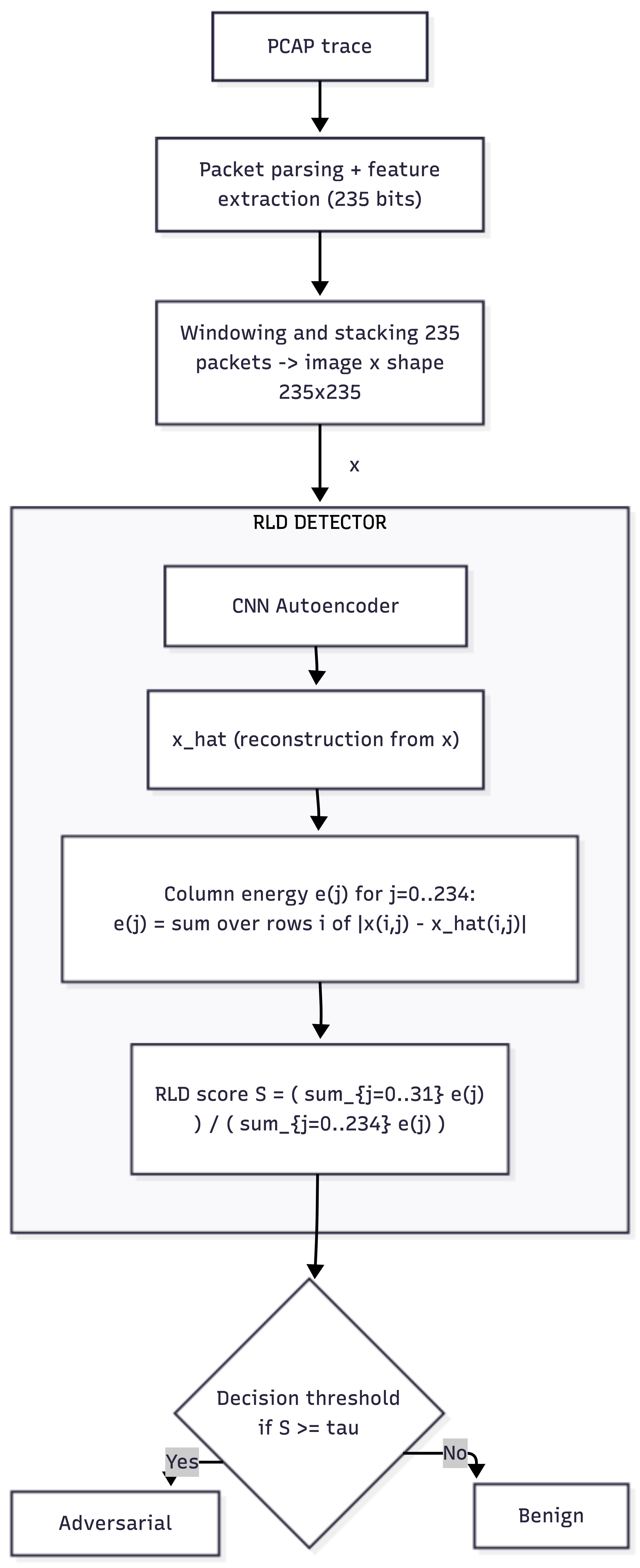}
    \caption{Residual Localisation Detector (RLD) pipeline.}
    \label{fig:rld}
\end{figure}

RLD measures how strongly the autoencoder's reconstruction residual is concentrated in the IAT feature region manipulated by the attacker. The pipeline is shown in Fig.~\ref{fig:rld}.

\subsubsection{Detection Procedure}
Let $x \in [0,1]^{H \times W}$ denote a packet image and $\hat{x}$ its reconstruction. The first $c = 32$ columns encode IAT bits. RLD computes:

\textbf{1) Absolute residual:} $R = |x - \hat{x}| \in \mathbb{R}^{H \times W}$

\textbf{2) Column-wise residual energy:} For each column $j$, $e_j = \sum_{i=0}^{H-1} R_{i,j}$

\textbf{3) Localisation score:}
\begin{equation}
S(x, \hat{x}) = \frac{E_{\text{IAT}}}{E_{\text{tot}} + \varepsilon} = \frac{\sum_{j=0}^{c-1} e_j}{\sum_{j=0}^{W-1} e_j + \varepsilon}
\label{eq:rld}
\end{equation}
where $\varepsilon > 0$ ensures numerical stability. By construction, $S \in [0,1]$: a score close to 1 indicates that nearly all reconstruction error is concentrated in the IAT columns. The detector flags a sample as adversarial if $S(x, \hat{x}) \geq \tau$.

\subsubsection{Design Rationale}
Focusing on the IAT columns directly reflects the threat model: PANDA's masked FGSM perturbs only IAT bits, so concentrating detection on this feature group exploits the attacker's known attack surface. Temporal behaviour is both a powerful signal for NIDS and relatively easy for an attacker to manipulate. Operating on residual localisation rather than raw reconstruction error makes the detector less sensitive to global variations in reconstruction quality. If the model reconstructs some header fields poorly on benign data, this increases $E_{\text{tot}}$ but does not concentrate error in the IAT region, keeping $S$ low. While our current design targets IAT due to PANDA's known perturbation strategy, the localisation principle generalises: a detector could monitor residual concentration across arbitrary feature groups to detect attacks that manipulate other packet fields.

\begin{figure}[t]
    \centering
    \subfloat[Clean traffic]{%
        \includegraphics[width=0.48\columnwidth]{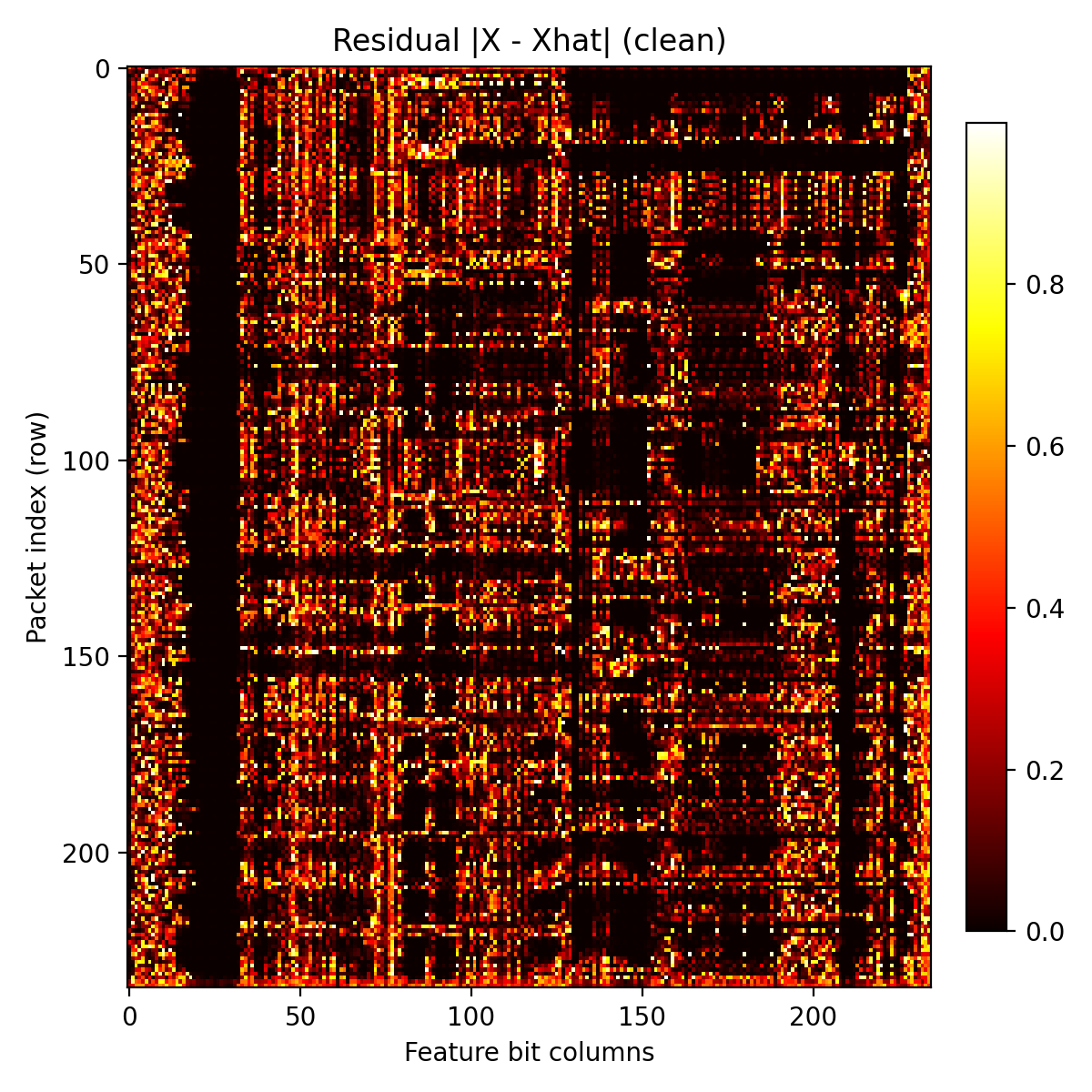}%
        \label{fig:residual_clean_RLD}%
    }\hfill
    \subfloat[Adversarial traffic]{%
        \includegraphics[width=0.48\columnwidth]{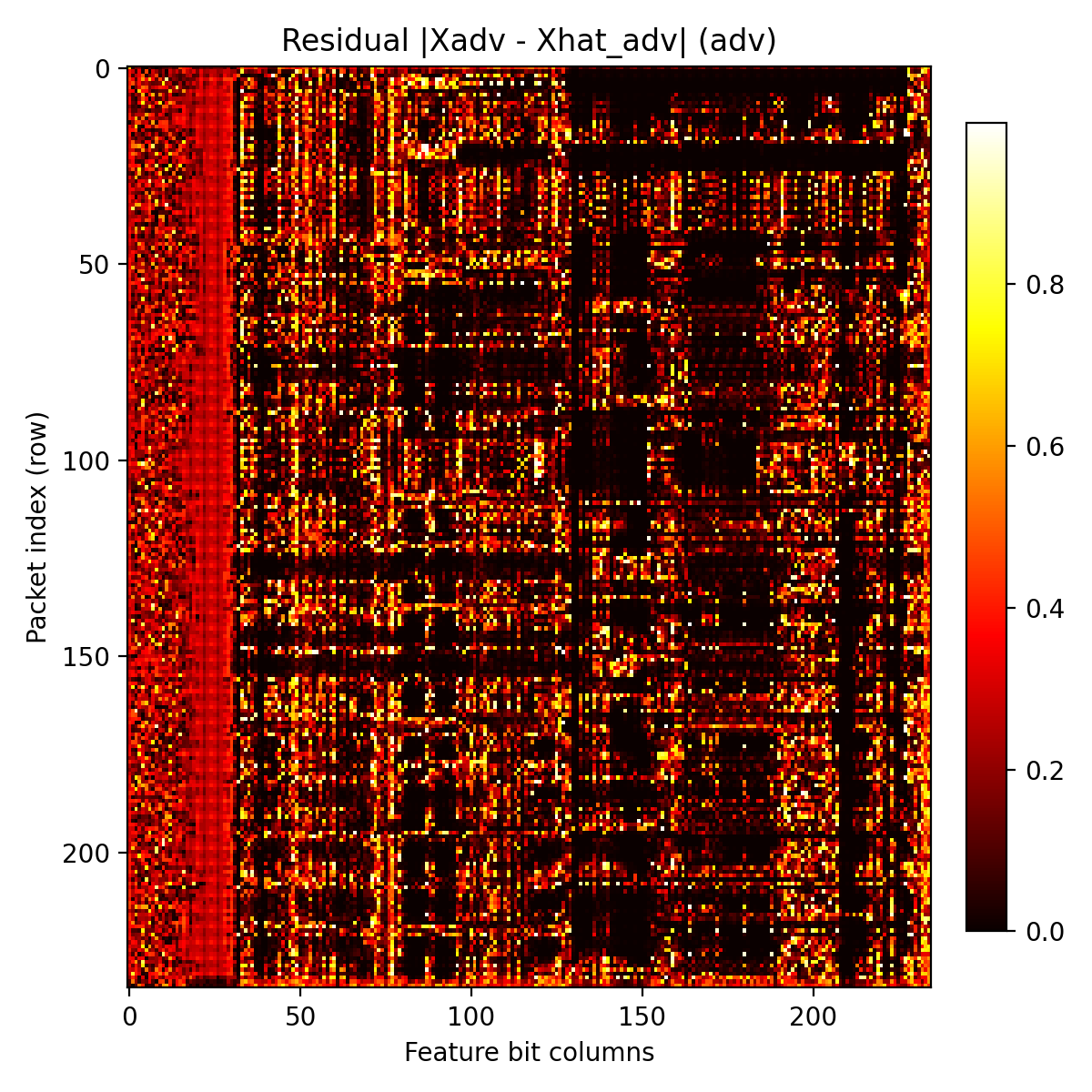}%
        \label{fig:residual_adv_RLD}%
    }
    \caption{Residual heatmaps for clean vs.\ adversarial traffic windows. Adversarial traffic shows strong residual localisation within the IAT feature region.}
    \label{fig:residual_comparison}
\end{figure}

Fig.~\ref{fig:residual_comparison} illustrates the key observation: adversarial samples exhibit residual energy strongly concentrated in the IAT columns, while clean traffic shows a diffuse pattern.

\subsection{Feature-Space Perturbation Consistency (FPC) Detector}

\begin{figure}[t]
    \centering
    \includegraphics[width=0.35\textwidth,height=0.65\textheight,keepaspectratio]{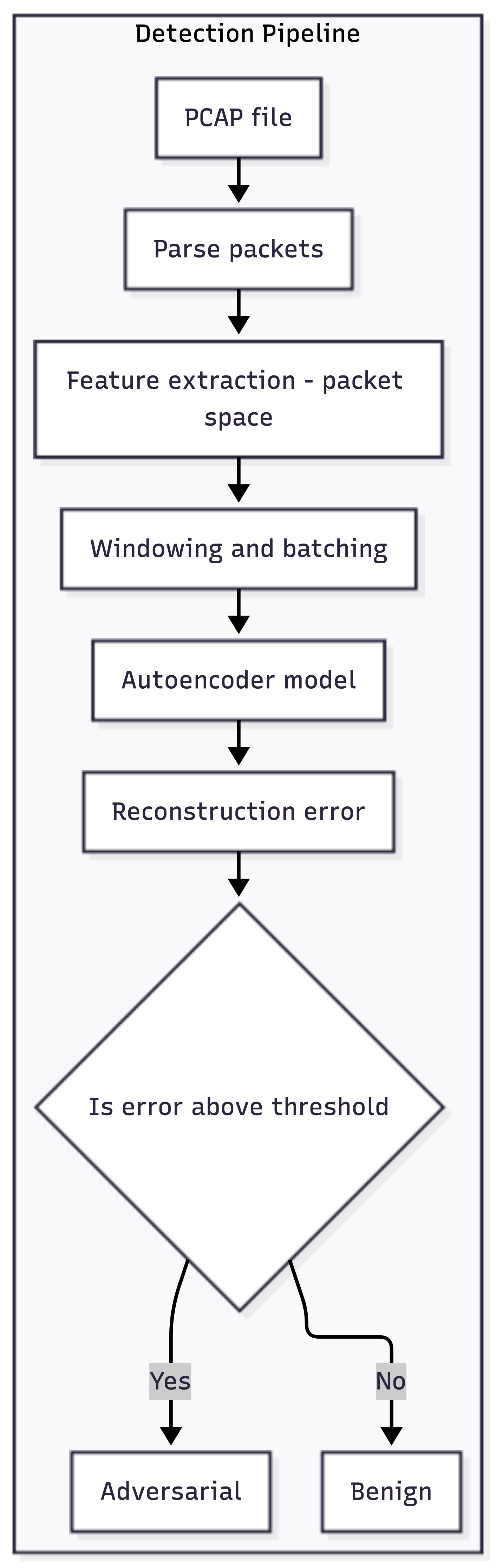}
    \caption{Feature-Space Perturbation Consistency (FPC) detector pipeline.}
    \label{fig:fpc}
\end{figure}

FPC leaves the image space entirely and operates directly on the 32-bit IAT encoding per packet. It trains a small fully-connected autoencoder on fixed-length windows of IAT bits and uses the per-window reconstruction error as an anomaly score. The pipeline is shown in Fig.~\ref{fig:fpc}.

\subsubsection{Detection Procedure}
Let $\{v_t\}_{t=1}^N$ be a sequence of IAT bit vectors with $v_t \in \{0,1\}^{32}$. For window length $W$, the sequence is segmented into windows $X_j \in \{0,1\}^{W \times 32}$, flattened to $x_j \in \{0,1\}^{d}$ with $d = W \cdot 32$. A shallow autoencoder $f_{\phi}: [0,1]^d \rightarrow [0,1]^d$ is trained on clean IAT windows using BCE loss. The encoder maps $x_j$ through two ReLU layers with dimensions $h$ and $h/2$; the decoder mirrors this with a final sigmoid layer. The anomaly score is:
\begin{equation}
R(X_j) = \frac{1}{d} \sum_{k=1}^{d} \mathrm{BCE}\bigl(x_{j,k}, \hat{x}_{j,k}\bigr)
\label{eq:fpc}
\end{equation}
Windows with $R(X_j) > \tau$ are flagged as adversarial, where $\tau$ is calibrated at a high percentile of clean reconstruction errors.

\subsubsection{Design Rationale}
Restricting input to 32 IAT bits per packet simplifies the model and targets the feature group most directly manipulated by PANDA's masked FGSM. This design choice is informed by knowledge of the attack strategy; however, the same approach could be extended to other packet feature groups (e.g., port numbers, frame lengths) by training separate autoencoders on their respective bit encodings. The default $W = 235$ reuses the same number of packets as PANDA's image representation for fair comparison. A shallow MLP autoencoder is chosen for simplicity and fast training on 1D bit vectors.

\subsection{Integration with NIDS}

The proposed detectors can be integrated with NIDS either sequentially or in parallel. In the sequential setup, traffic is first passed through the NIDS; malicious traffic is blocked, while traffic classified as benign is forwarded to the detectors, which may flag it as adversarial. In the parallel setup, traffic passes through both NIDS and detectors simultaneously; traffic is considered benign only if both modules agree.

\section{Evaluation}

\subsection{Experimental Setup}

\subsubsection{Surrogate Autoencoder}
All adversarial examples are generated against the CNN autoencoder from the PANDA framework~\cite{swain2024panda}. The encoder consists of two convolutional layers with kernel size 3, ReLU activations, and $2 \times 2$ max-pooling, compressing from 1 to 32 to 64 channels. The decoder mirrors this with transposed convolutions and a final sigmoid layer. Training uses RMSE loss on benign traffic only.

\subsubsection{Limitation of Saved Adversarial PCAPs}
\label{sec:pcap_degradation}
\begin{figure}[t]
    \centering
    \includegraphics[width=0.8\columnwidth]{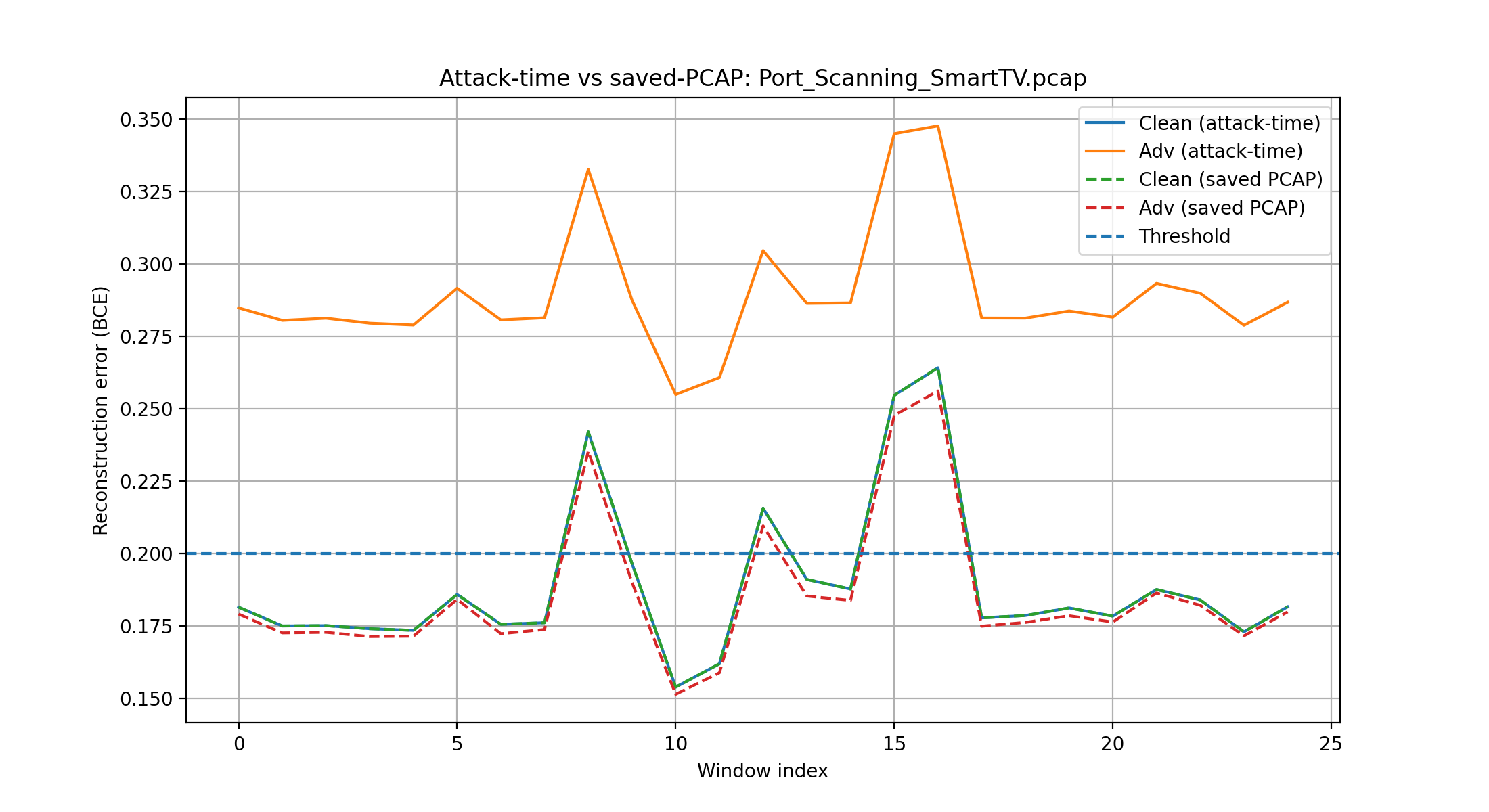}
    \caption{Effectiveness of adversarial perturbation on in-memory tensors vs.\ saved PCAPs. Adv (in-memory) changes the reconstruction error, whereas Adv (saved PCAP) reverts to near the clean value.}
    \label{fig:in_memory_vs_saved}
\end{figure}

We observed that saving adversarial examples as PCAP files and re-parsing them degrades or erases their adversarial characteristics. Inter-arrival times must be rounded to integer microseconds and timestamps must move forward, partially undoing the carefully crafted perturbations. Fig.~\ref{fig:in_memory_vs_saved} shows this effect: adversarial in-memory tensors clearly shift the reconstruction error, while saved-PCAP versions revert to nearly original malicious scores. RLD uses in-memory adversarial tensors, while FPC operates on re-parsed PCAPs.

\subsubsection{Adversarial Traffic Generation}
FGSM attacks are sensitive to the choice of $\epsilon$. As shown in Fig.~\ref{fig:epsilon_comparison} (Appendix), evasion is optimal at $\epsilon = 0.5$, which we fix for all experiments. For each malicious window, the FGSM attack produces an adversarial version in memory. RLD operates directly on these in-memory adversarial tensors. For FPC, the adversarial tensors are converted back to PCAP files and re-parsed to extract IAT features, which partially degrades the adversarial perturbations as described in Section~\ref{sec:pcap_degradation}.

\subsubsection{Evaluation Metrics}
We evaluate at the window level using standard binary classification metrics: true negative rate (TNR), true positive rate (TPR), precision, recall, and F1-score. TNR measures how often the detector leaves benign windows unflagged; TPR measures how often adversarial windows are correctly flagged.

\subsection{RLD Detector Results}

\begin{figure}[t]
    \centering
    \includegraphics[width=0.8\columnwidth]{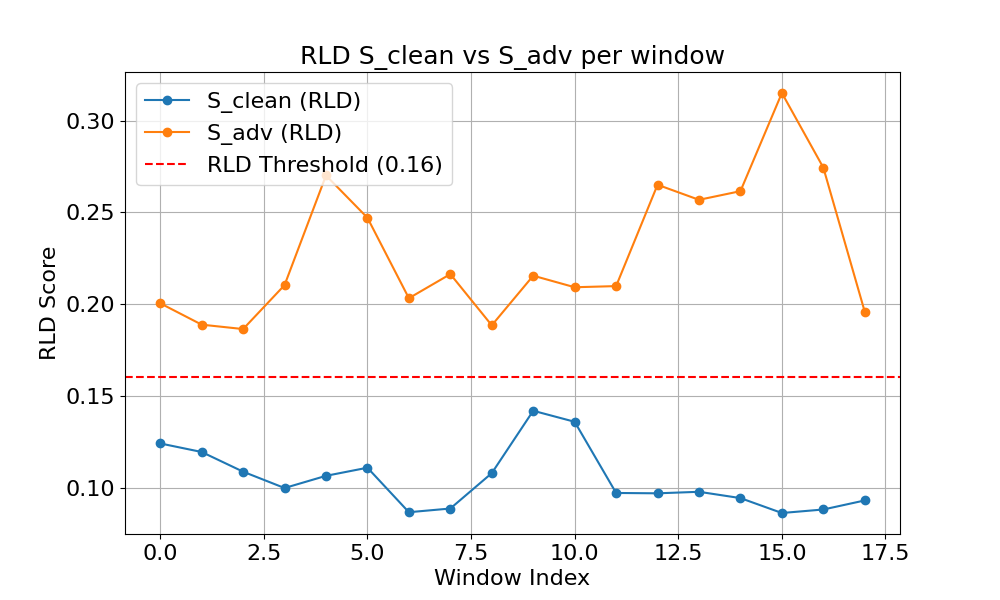}
    \caption{RLD scores for clean vs.\ adversarial windows. Clear separation is observed with a global threshold.}
    \label{fig:rld_results}
\end{figure}

The RLD detector uses a global threshold $\tau = 0.160$, flagging windows with $S \geq \tau$ as adversarial. Fig.~\ref{fig:rld_results} and Table~\ref{table:rld} summarise the results. Across 78,116 windows, RLD achieves TNR $\approx$ 0.9999 and TPR $\approx$ 0.9972, with only 2 false positives and 111 false negatives. The detector shows strong discrimination across diverse traffic scenarios.

\begin{table}[t]
\centering
\caption{RLD window-level evaluation over all pcap traces ($\tau = 0.160$).}
\label{table:rld}
\begin{tabular}{lr}
\toprule
\textbf{Quantity} & \textbf{Value} \\
\midrule
Total windows & 78,116 \\
Clean / Adversarial windows & 39,058 / 39,058 \\
True Negative / False Positive & 39,056 / 2 \\
False Negative / True Positive & 111 / 38,947 \\
TNR & 0.9999 \\
Recall (TPR) & 0.9972 \\
Precision / F1-score & 0.9999 / 0.9986 \\
\bottomrule
\end{tabular}
\end{table}

\textbf{Sensitivity analysis.} On IoT cameras, clean RLD scores average $\approx 0.037$ while adversarial scores reach $\approx 0.32$, leaving a large gap around $\tau = 0.160$. For ARP spoofing, the gap narrows (clean $\approx 0.12$, adversarial $\approx 0.19$) but remains sufficient.

\textbf{Failure cases.} The most visible weaknesses occur for Port Scanning and Service Detection on the Lenovo Bulb device, where TPR drops to 0.28 and 0.24 respectively. These cases suggest RLD may struggle when adversarial traffic reconstruction resembles benign patterns.

\subsection{FPC Detector Results}

\begin{figure}[t]
    \centering
    \includegraphics[width=0.8\columnwidth]{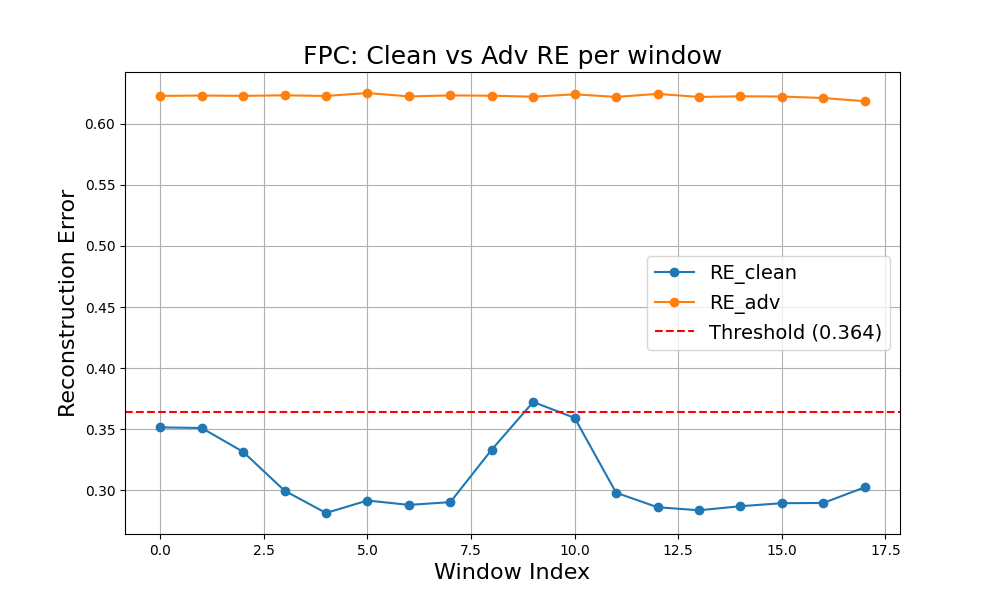}
    \caption{FPC reconstruction errors for clean vs.\ adversarial traffic. Adversarial values consistently exceed the threshold.}
    \label{fig:fpc_results}
\end{figure}

Unlike RLD, FPC operates on saved adversarial PCAPs (which partially lose adversarial characteristics). Despite this harder setting, FPC achieves TNR $\approx$ 0.999 and TPR $\approx$ 0.986, as shown in Fig.~\ref{fig:fpc_results} and Table~\ref{table:fpc}. The threshold $\tau = 0.36$ is set at the 99th percentile of clean reconstruction errors.

\begin{table}[t]
\centering
\caption{FPC window-level evaluation over all pcap traces ($W=235$, $\tau = 0.36$).}
\label{table:fpc}
\begin{tabular}{lr}
\toprule
\textbf{Quantity} & \textbf{Value} \\
\midrule
Total windows (clean / adversarial) & 1,342 / 1,342 \\
True Negative / False Positive & 1,341 / 1 \\
False Negative / True Positive & 19 / 1,323 \\
TNR & 0.9993 \\
TPR (Recall) & 0.986 \\
Precision / F1-score & 0.9992 / 0.9926 \\
\bottomrule
\end{tabular}
\end{table}

\subsection{Comparative Summary}

Table~\ref{table:comparison} compares both detectors. RLD provides near-perfect detection with minimal false positives in image space. FPC, despite operating on degraded saved PCAPs, still achieves strong performance in packet-feature space. Each detector exploits a different signal: RLD measures residual localisation, and FPC measures feature-space anomaly.

\begin{table}[t]
\centering
\caption{Comparison of the two proposed detectors.}
\label{table:comparison}
\begin{tabular}{lcc}
\toprule
\textbf{Metric} & \textbf{RLD} & \textbf{FPC} \\
\midrule
Operating space & Image & Packet \\
Threshold $\tau$ & 0.160 & 0.36 \\
Total windows & 78,116 & 2,684 \\
TNR & 0.9999 & 0.9993 \\
TPR & 0.9972 & 0.986 \\
Precision & 0.9999 & 0.9992 \\
F1-score & 0.9986 & 0.9926 \\
\bottomrule
\end{tabular}
\end{table}

\section{Discussion and Limitations}

While both detectors achieve strong performance, several limitations should be acknowledged.

\textbf{Threshold generalisation.} All thresholds were calibrated on traffic from the UQ-IoT dataset. Deployment in environments with more diverse traffic patterns would likely require recalibration. The same global threshold may not generalise to completely unseen networks without adjustment.

\textbf{Restricted feature space.} The current detectors are designed for PANDA's IAT-only perturbation model, exploiting the fact that PANDA's masked FGSM perturbs only the IAT columns. This design is informed by knowledge of the attack strategy: both detectors focus on IAT because that is the known attack surface. An attacker who shifts perturbations to other header fields or payload features, while keeping IAT patterns close to benign, could potentially evade the detectors. Investigating how RLD and FPC perform when monitoring multiple packet feature groups—such as port numbers, IP addresses, and frame lengths—beyond IAT alone is an important direction for making these detectors robust against a wider range of adversarial strategies.

\textbf{Adaptive adversaries.} Our evaluation assumes the attacker is unaware of the detectors. An adaptive attacker who optimises directly against the detectors could degrade performance. Designing and evaluating detectors against adaptive attacks is left for future work.

\textbf{PCAP round-trip degradation.} We observed that PANDA-style perturbations lose efficacy when adversarial traffic is saved as PCAP files and re-parsed. While this limits the practical threat of the current PANDA implementation, future attacks may incorporate more robust perturbation strategies that survive PCAP round-trips. The FPC detector's strong performance even on degraded adversarial PCAPs is encouraging in this regard.

\textbf{Single attack family.} Our evaluation considers only FGSM-based attacks from the PANDA family. More sophisticated attacks (iterative, adaptive, or based on different gradient methods) may produce different adversarial signatures and warrant further investigation.

\section{Conclusion}

This paper addressed the problem of detecting adversarial evasion attacks against autoencoder-based NIDS. The PANDA framework demonstrates that network packets can be converted into an invertible image representation, enabling gradient-based FGSM attacks that evade detection while preserving malicious functionality. We proposed two complementary detectors: RLD tracks spatial concentration of reconstruction errors in the IAT region in image space, and FPC operates directly on packet-level IAT features in packet-feature space.

Both detectors achieve near-perfect detection performance (F1 $\geq$ 0.99) across multiple IoT devices and attack types in the UQ-IoT dataset. RLD provides near-perfect detection with minimal false positives in image space, and FPC demonstrates strong performance even on degraded adversarial PCAPs.

These results support three conclusions. First, PANDA-style adversarial attacks are feasible against realistic IoT traffic and can bypass naive threshold-based NIDS. Second, carefully designed detectors that combine reconstruction information with perturbation-based consistency checks can reliably detect such attacks without high false positive rates. Third, no single detector is universally sufficient: RLD and FPC each have distinct strengths and weaknesses, and combining these views is a promising direction for future work.

Future work should investigate detector generalisation to multiple packet feature groups beyond IAT, broaden the threat model to perturbations targeting other header fields, investigate adaptive attackers that optimise against detectors, develop PCAP-robust adversarial attacks that survive packet-level round-trip transformations, explore co-training strategies for NIDS and detectors, and develop common datasets and evaluation protocols for fair comparison.

\bibliographystyle{ACM-ReferenceFormat}
\bibliography{references}

\appendix
\section{FGSM Sensitivity to \texorpdfstring{$\epsilon$}{Epsilon}}

\begin{figure}[h]
\centering
\subfloat[$\epsilon = 0.1$]{%
  \includegraphics[width=0.18\textwidth]{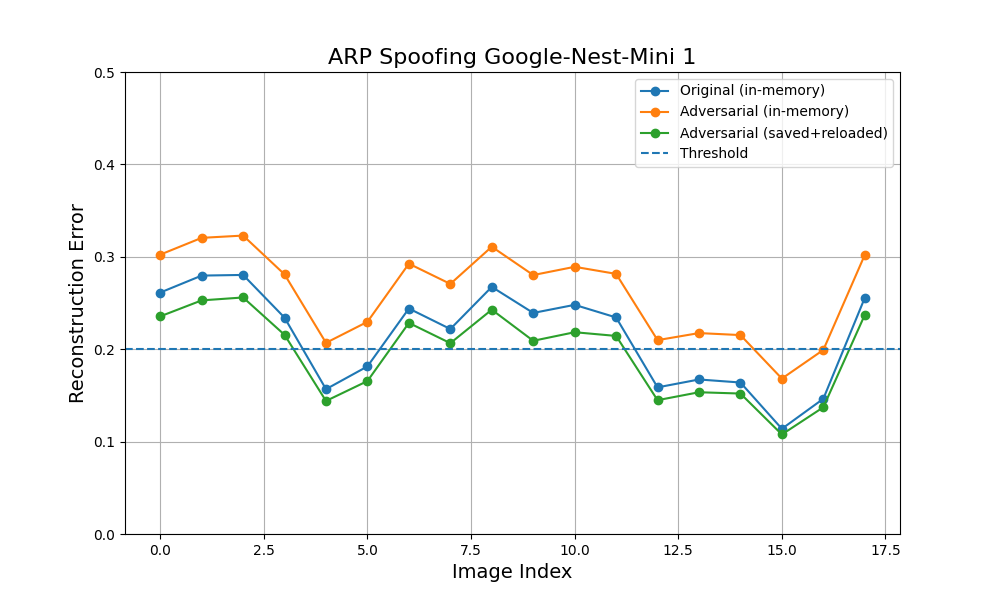}%
}\hfill
\subfloat[$\epsilon = 0.3$]{%
  \includegraphics[width=0.18\textwidth]{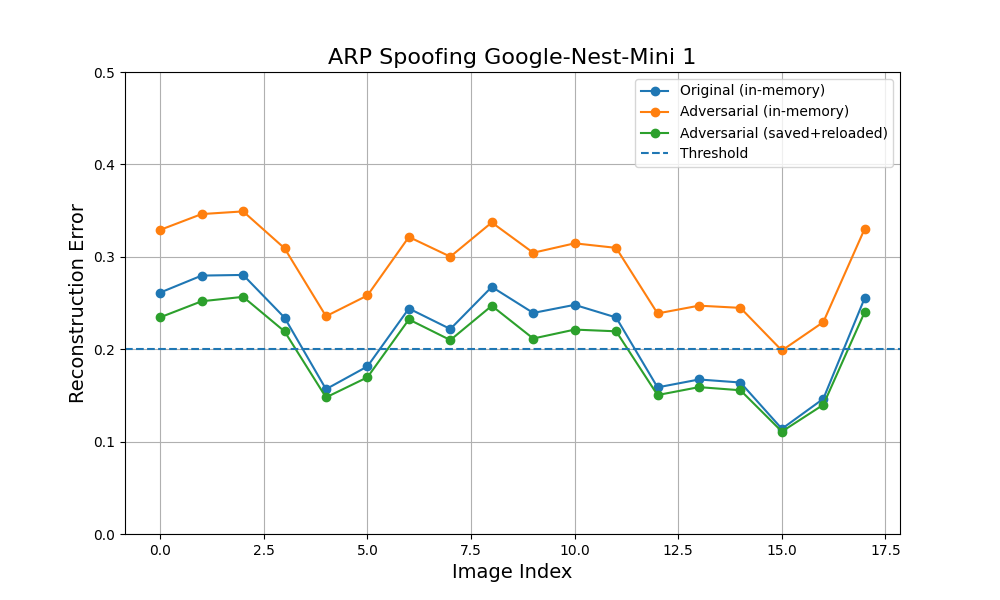}%
}\hfill
\subfloat[$\epsilon = 0.5$]{%
  \includegraphics[width=0.18\textwidth]{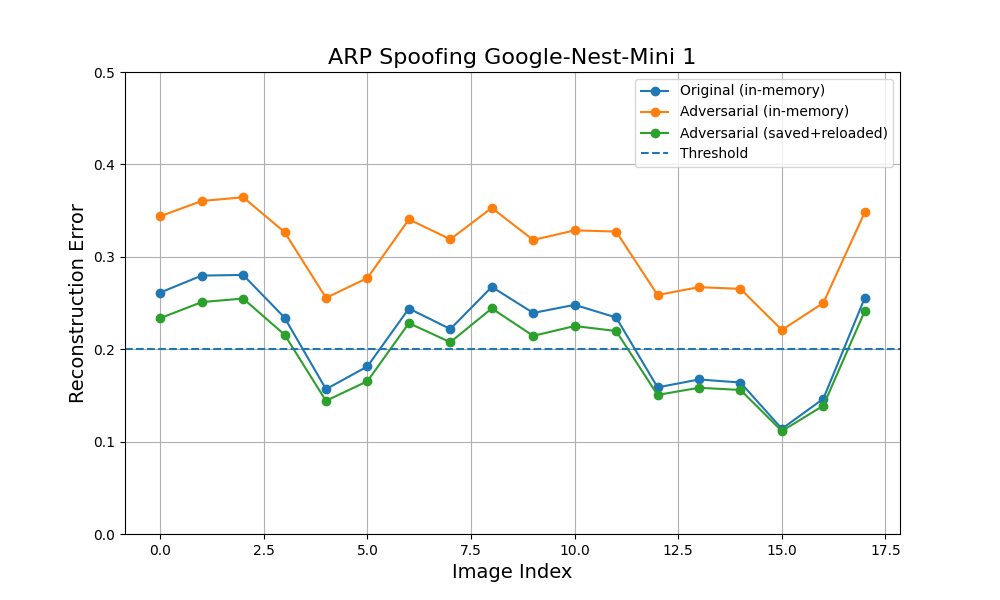}%
}\hfill
\subfloat[$\epsilon = 0.7$]{%
  \includegraphics[width=0.18\textwidth]{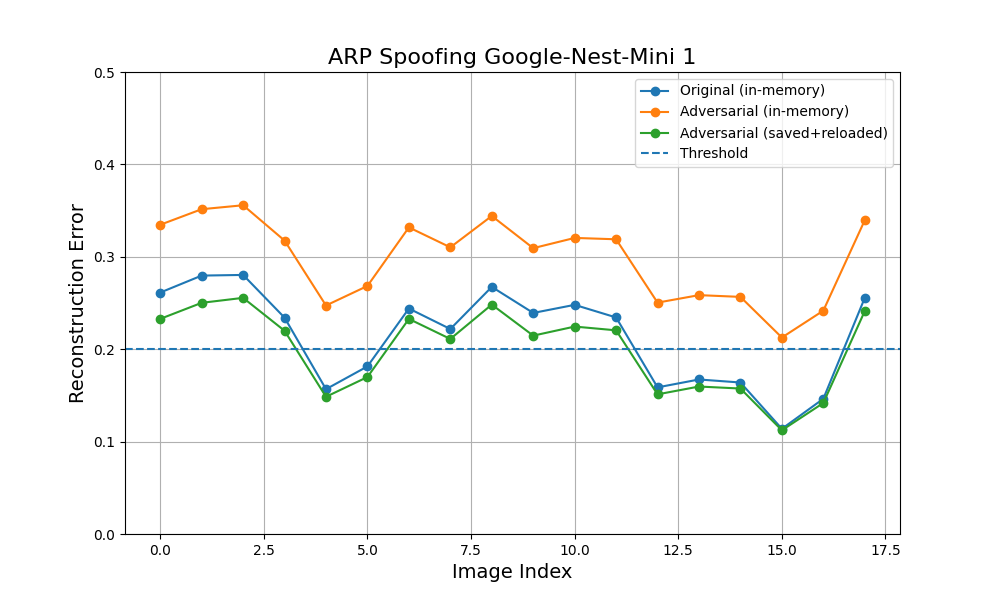}%
}\hfill
\subfloat[$\epsilon = 0.9$]{%
  \includegraphics[width=0.18\textwidth]{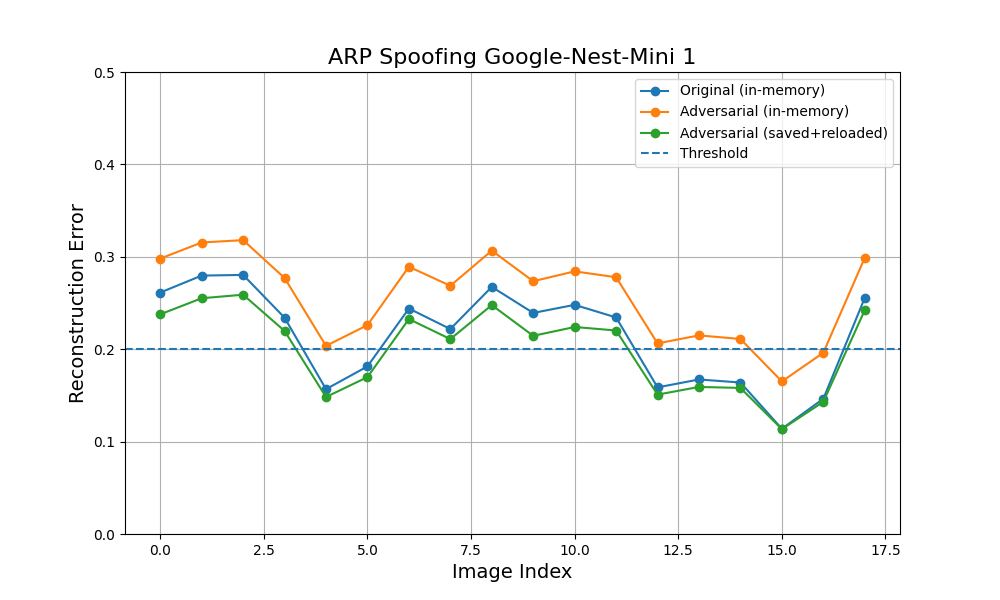}%
}
\caption{FGSM sensitivity to $\epsilon$. Reconstruction error increases from $\epsilon = 0.1$ to $0.5$, then decreases. Optimal: $\epsilon = 0.5$.}
\label{fig:epsilon_comparison}
\end{figure}

\end{document}